\documentclass[11pt,twoside]{article}

\usepackage{asp2006}
\usepackage{epsf}
\usepackage{epsfig}
\usepackage{lscape}
\usepackage{natbib}
\def\lsi{{LS~I~$+61^\circ\,303$}}

\begin{document}
\title{VSOP and stellar sources - the case of LS I $+61^\circ303$}   
\author{S.M. Dougherty}   
\affil{National Research Council Herzberg Institute for Astrophysics\\
Dominion Radio Astrophysical Observatory}    

\begin{abstract} 
Space-VLBI observations of stellar sources represent a challenge since
there are few sources with sufficiently high brightness temperature
for detection on space-ground baselines. X-ray binaries (XRB) are
among the few types of stellar radio sources that can be
detected. Observations of the unusual X-ray and $\gamma$-ray binary
system \lsi~ obtained with the HALCA satellite and a 20-element ground
array are described.  The data in this 48-hour experiment represent
some of the best quality VLBI observations of \lsi. No fringes were
detected on HALCA baselines. 10-minute snapshot images were produced
from the global ground array data and reveal an expansion velocity of
$800$~km~s$^{-1}$. Some of these image data reveal hints of more
extended emission but high-SNR closure phase data do not support
relativistic outflow in the plane-of-the-sky in \lsi. The largest
closure phase rates are consistent with an outflow of
$\sim1000$~km~s$^{-1}$ as deduced from the image data. The closure
phases also show no evidence of structure variation on size scales
greater than $\sim10$~mas.  A number of issues related to VSOP2
observations of stellar radio sources are raised.
\end{abstract}


\section{Introduction}

A challenge of observational astrophysics with radio
interferometers is the fundamental limitation the observing array
places on brightness temperature sensitivity. 
The sensitivity of the individual telescope elements in the array
determines the flux density limit. The HALCA spacecraft had a
$7\sigma$ sensitivity in a few minutes to a ground-based telescope of
$\sim100$~mJy, and so for an array with $B_{\rm max}\sim10,000$~km (a
resolution of $\sim1$~mas at 5~GHz) this implies a brightness
temperature limit of $T_B\sim 10^{10}$~K. Consequently the types of
stellar sources that could be detected on baselines to HALCA was
limited to masers, pulsars, and some XRBs. There are undoubtedly
others types of sources that could be detected, but they will exhibit
either outburst phenomena or coherent emission processes.

The subjects of maser and pulsar observations are well covered by
other papers presented at the meeting and the previous VSOP
symposium. In this paper, the specific case of VSOP observations of
the high mass XRB \lsi~will be described. Special attention will be
given to some properties of the radio emission in \lsi~as a means to
reveal some of the challenges that need to be addressed for future
space-VLBI observations of stellar sources.

\section{What is LS I $+61^{\circ}303$?}
The luminous early-type star \lsi~was discovered as a radio source by
\citet{Gregory:1978}, who proposed an association with the COS-B
$\gamma$-ray source 2CG 135+01 \citep{Hermsen:1977}. The radio
emission was shown to have regular, non-thermal outbursts with a
period of 26.51 days \citep{Taylor:1984}; the flux density rises from
a quiescent level of around 20~mJy to a peak of up to 300 mJy over
$\sim2$~days, and then decays over $\sim8$~days to its quiescent
level. The period is related to the highly eccentric orbit of a
compact object around a rapidly rotating B0e star
\citep{Hutchings:1981,Grundstrom:2007}.  The 26.5-day periodic
behavior is also manifest at other wavelength regions: IR
\citep{Paredes:1994}, X-ray \citep{Paredes:1997} and, most recently
$\gamma$-ray energies \citep{Albert:2006, Maier:2007}.

Variable TeV emission has been detected in three other high-mass XRB
systems. 
Together they form a group of objects that have been referred to as
Binary TeV sources (BTVs).  PSR B1259-63 is a 3.4-yr period massive
binary system with a pulsar in an eccentric orbit around a B0e star
\citep{Johnston:1992}. The TeV and X-ray emission is interpreted as
inverse-Compton (IC) emission due to relativistic electrons
accelerated in a wind-collision between the pulsar wind and that of
the massive companion \citep[e.g.][]{Tavani:1997}. Cygnus X-1 is a
well-established micro-quasar system with clearly observed
relativistic jets \citep[e.g.][]{Stirling:2001, Fender:2006}. The
situation in both LS~5039 and \lsi~is far from clear. In LS5039, there
is evidence of relativistic jets but the picture is not completely
clear \citep{Ribo:2008}.  In \lsi, both micro-quasar
\citep[e.g.][]{Massi:2004} and pulsar-wind nebulae \citep{Dhawan:2006}
models have been advanced, with the merits of each scenario discussed
by \citet[][and references therein]{Romero:2007}. The remainder of
this paper will discuss the insights revealed by VSOP observations of
\lsi~and how they may inform this discussion.

\section{Previous VLBI observations}
\label{sec:prevobs}
VLBI observations of \lsi~have been completed by a number of different
{groups. Observations with the EVN near peak flux indicate a low
expansion velocity of $\sim600$~km~s$^{-1}$
\citep{Taylor:1992}. However, an expansion velocity of
$\sim18,000$~km~s$^{-1}$ was measured during a mini-outburst observed
during the main outburst \citep{Peracaula:1998}.  More recently,
evidence for relativistic expansion has been presented from EVN and
MERLIN observations. \citet{Massi:2001} observed an extension of the
central source to the SE. The lack of a corresponding symmetric
feature to the NW led to an interpretation of a Doppler amplified
outflow close to the line-of-sight with an inferred ejection speed of
$0.4$c. A more recent MERLIN observation suggests a similar structure,
though with a double-sided jet that extends up to 200 AU ($\sim70$~mas)
from the central source and which precesses significantly in 24 hours
\citep{Massi:2004}. This is interpreted as due to a jet with an
outflow speed of 0.6c. Clearly, these observations do not present a
consistent picture of the dynamical evolution of the radio emission
arising in \lsi.

\begin{figure}[!t]
\plotfiddle{figure2.ps}{4cm}{-90}{29}{29}{-19}{140}
\plotfiddle{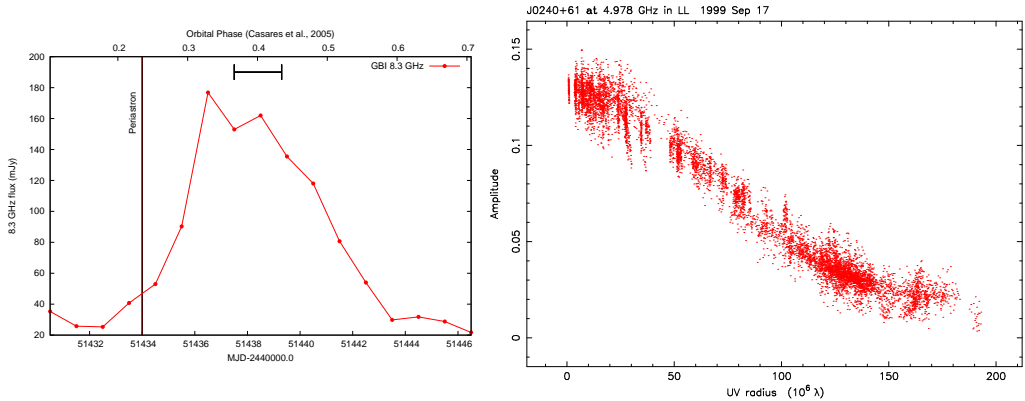}{0cm}{0}{50}{50}{-220}{-13}
\caption[]{Left: The epoch of VSOP observations (horizontal bar)
relative to an outburst in \lsi~as observed by the GBI. The
orbital phase from \citet{Casares:2005} is shown, and the phase of
periastron. Right: Visibility distribution of \lsi~over 1 hour.  The
source is clearly resolved.
\label{fig:outburst}}
\end{figure}
\section{VSOP observations of LS I $+61^{\circ}303$}
Observations over 48 hours were obtained on 1999 September 17 at 5 GHz
with HALCA and a 20-element ground array including some of the largest
telescopes available (the phased-VLA, Effelsberg, Jodrell Bank). The
goal was to image \lsi~with sub-milliarcsecond resoltion to observe
structural changes as the source evolved during the 2 days of
outburst. Though VSOP observations pre-date some of the observations
described in Sec.~\ref{sec:prevobs}, they represent perhaps the
largest global array assembled to observe \lsi, and such high-SNR data
is arguably the best available today for this source. They
have been described previously by \citet{Taylor:2000}, but more recent
analysis is included here.

Unfortunately, the observations started approximately 60 hours after
the onset of outburst (Fig~\ref{fig:outburst} - left), and data was
collected during decline.  Furthermore, fringes on baselines to the
HALCA spacecraft were not detected. The visibility distribution from
part of the observation is shown in Fig~\ref{fig:outburst} (Right). This 
plots demonstrates that at the spatial frequencies sampled by HALCA baselines
($>200$M$\lambda$), the correlated flux ($\sim20$~mJy or less) is 
too low for direct fringe detection (see comments in
Sec.~\ref{sec:vsop2}).

The ground-telescope array data alone provides a very high-quality
VLBI observation of \lsi~during the decline phase of an outburst. Standard
phase-referencing techniques were used in the ground observations,
enabling antenna gain calibration using the bright quasar J0228+67 at
a few degrees distance every 20 minutes.  Snapshot images every 10
minute were generated, with two examples from 112 total images shown
in Fig.~\ref{fig:snapshots}.


\begin{figure}[!ht]
\plotfiddle{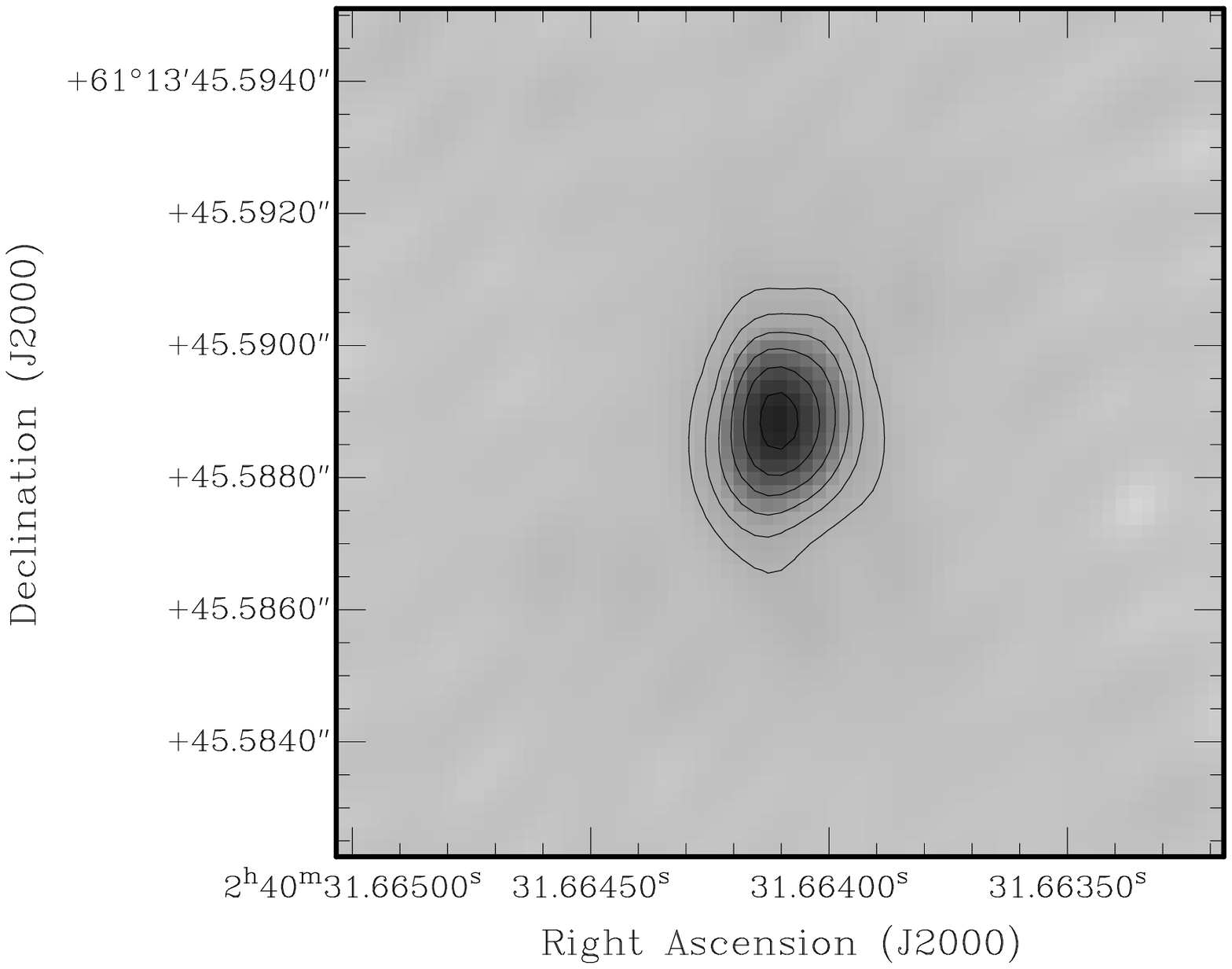}{5cm}{0}{43}{43}{-50}{-96}
\plotfiddle{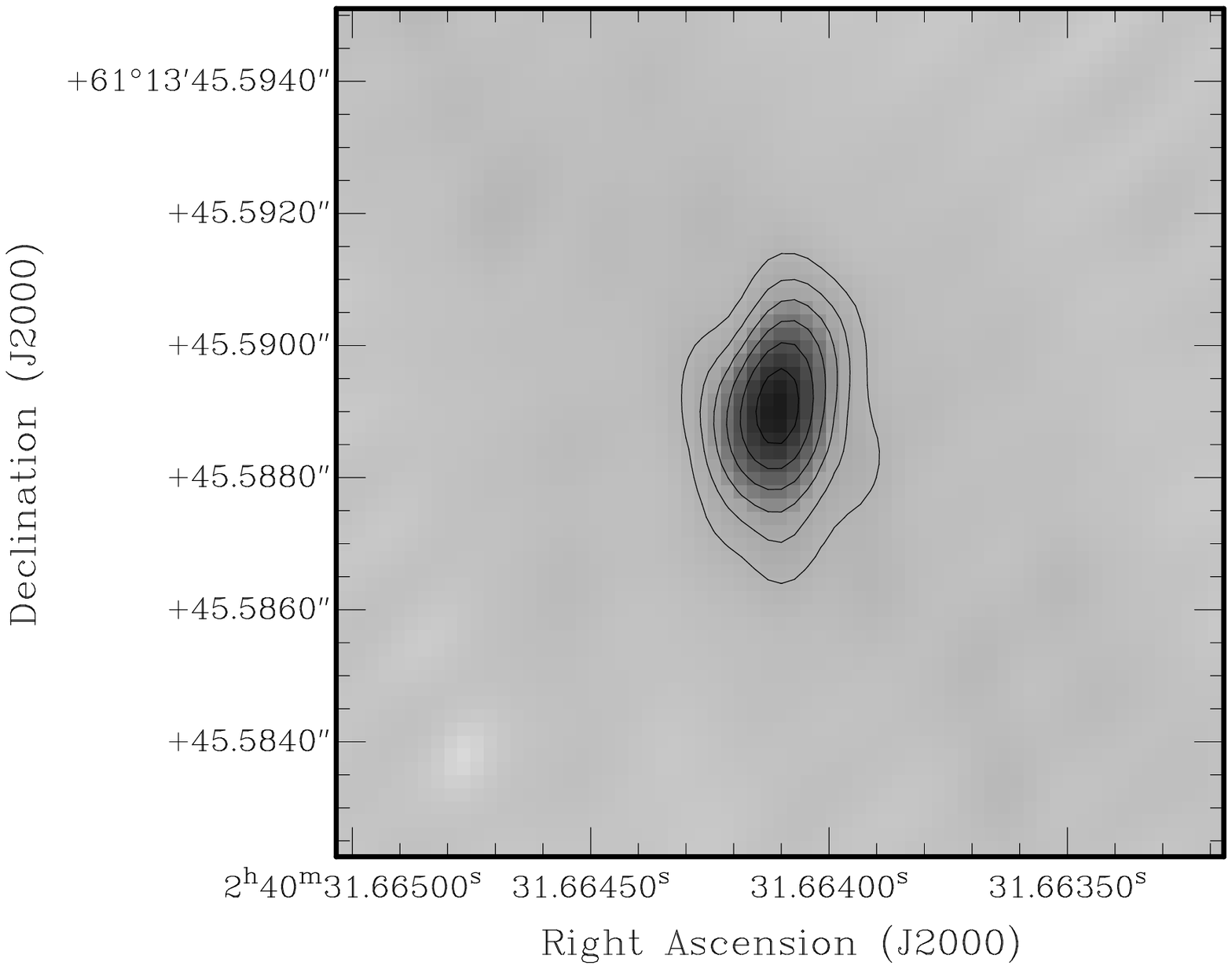}{0cm}{0}{43}{43}{-210}{-72}
\caption[]{Two 10-minute images from the 113 in the
observation. Contours are $8, 12, 18\dots80\%$ of peak. Frame~21
(left) and Frame 79 (right) are typical examples.  Visibility modelling of data from each
day reveals expansion of $\sim0.3$~mas over 24 hours.
\label{fig:snapshots}}
\end{figure}


\begin{figure}[!ht]
\plotone{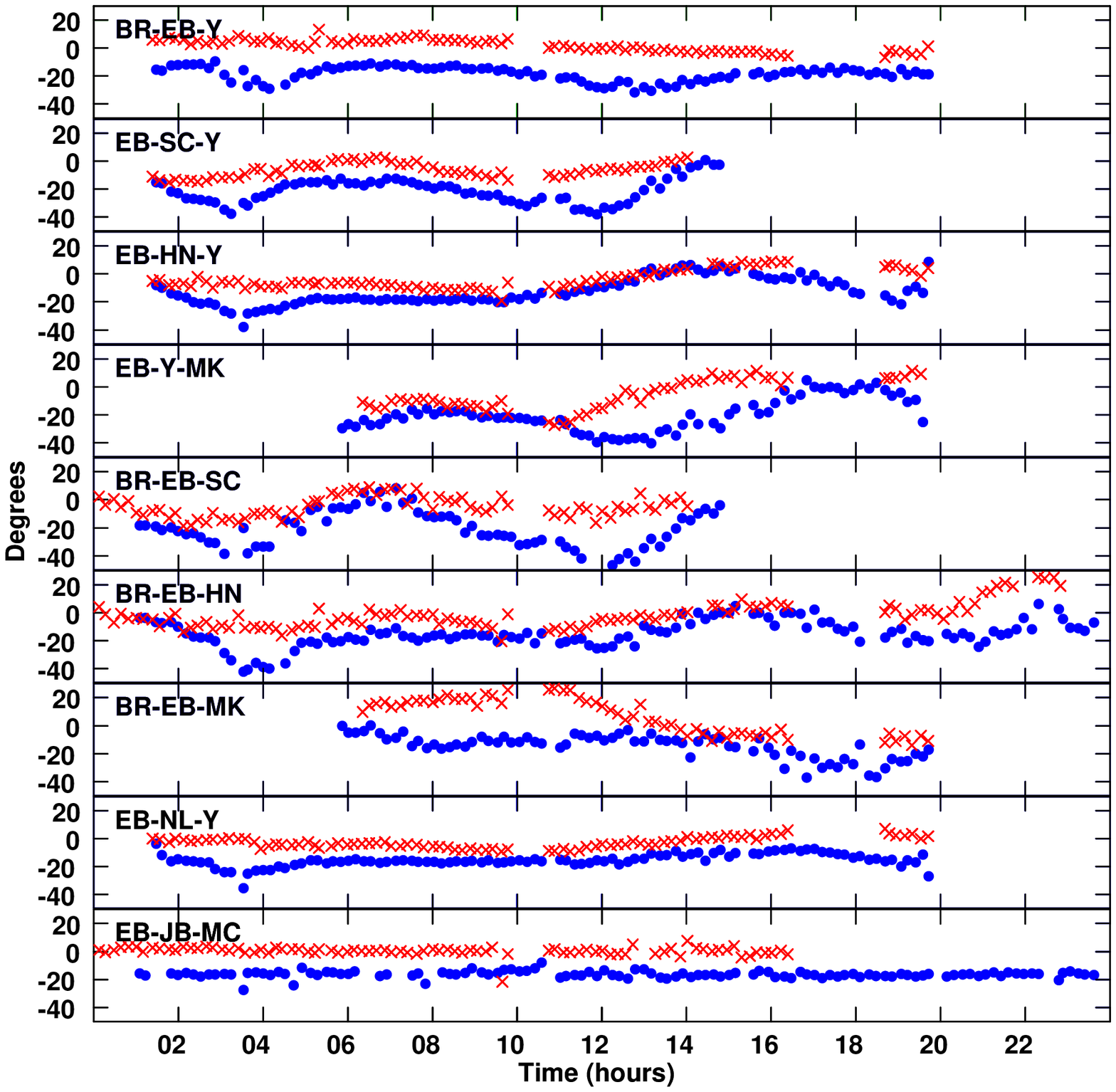}
\caption[]{Closure phase every 10 minutes from September 16 (dots) and
17 (crosses) (Vivek Dhawan, priv. comm.) The closure phase from each
day is offset by $20^\circ$ for clarity . A number of different
three antenna groups are shown, all with very high SNR data.
\label{fig:closure}}
\end{figure}

The image data show the source has a maximum extent of $\sim4$~mas,
similar to previous observations taken during quiescence.  Comparison
of visibility data on each of the two days shows evidence for a
$\sim0.3$~mas angular expansion over 24 hours, corresponding to an
expansion of 0.6~AU at a distance of 2~kpc \citep{Frail:1991}, and an
inferred expansion velocity of $\sim800$~km~s$^{-1}$.

A feature seen in some of these images is low surface brightness
extensions to the central emission region. These extensions are
roughly symmetric about the central source and have been interpreted
as emission from high-velocity outburst ejecta.  If they are
associated with the outburst the inferred expansion velocity is
$42,000$km~s$^{-1}$, and broadly consistent with expansion
deduced from MERLIN and EVN observations by Massi et al.

However, interpretation of the extensions in some of the images is
hindered by a) the inability to follow the features from one 10-minute
snapshot to the next and b) the symmetric nature of the features. This
led us to consider if these were artefacts of calibration errors. A
robust technique for eliminating calibration artefacts is to examine
closure quantities, particularly phase where any time variation in
closure phase is due {\em only} to intrinsic properties of the radio
source. This is particularly useful with high-SNR data since even
small closure phase variations can be readily detected.

Closure phase for a number of antenna triangles is shown in
Fig.~\ref{fig:closure} and several things can be deduced
readily. Firstly, on 10-mas size scales represented by the EB-JB-MC
triangle (bottom frame), the closure phase is identically zero. This
implies \lsi~is a point source on these angular scales with no
variable structure during this observation. The next important feature
is that the largest changes in closure phase are $\sim20^\circ$ over a
few hours.  What does this imply?  $20^\circ$ is approximately 1/20 of
the beam so for a 1~mas beam this corresponds to an asymmetric
structure changes of $\sim50\mu$as. Over a few hours, this corresponds
to an expansion of $\sim1000$km~s$^{-1}$, consistent with the
visibility modelling analysis. Most importantly, at 2~kpc an expansion
in the plane-of-the-sky at $1c$ would result in a closure phase rate
of $180^\circ$ in 8 minutes, which would be easily seen, especially in
such high-SNR data. Though the phase rate is dependent on the degree
of asymmetry of the structure change, there is no evidence of
relativistic expansion in the closure phase. This leads to the
conclusion that the extensions are an artefact of calibration error.

%
%
%

\section{VSOP2 and stellar radio astrophysics}
\label{sec:vsop2}
Stellar observations with VSOP were particularly challenging due to
the lack of the necessary sensitivity on HALCA to directly detect
fringes.  However, the observations of \lsi~provide some useful
insight for planning stellar radio observations with future space-VLBI
missions such as VSOP2.

It is important that sensitivity is maximized.  This
implies the use of the most sensitive ground-based telescopes, not
only in terms of collecting area (e.g. phased-VLA, Effelsberg, GBT
etc), but also the observing frequency.  This suggests 8.6 GHz will be
the most suitable frequency since it offers an ideal combination of
sensitivity on the ground and, for non-thermal sources, a higher flux
than at the higher frequency observing bands specified for
VSOP2.  Ideally, the most sensitive 8.6-GHz system possible should be
available on the spacecraft.

An imperative for stellar observations in future missions is a
source switching or ``phase-referencing'' capability with the
spacecraft. This ability was not available with HALCA. First, it is
important to be able to involve the spacecraft in fringe-finding
observations in order to establish the clock and orbit geometry
offsets. If a source is sufficiently bright, phase-referencing
observations can be done to utilise self-calibration to solve for time
evolution of these parameters and the gain phase of the
spacecraft. Further more, this provides an ability to improve orbit
definition through astrometry.

The VSOP observations of \lsi~were planned for a predicted time of
outburst. Unfortunately, the onset of outburst was missed by 60
hours. Observations during outburst could only have been assured
through a dynamic scheduling capability.  Certainly if transient
science is a key project for VSOP2 then a dynamic scheduling
capability that permits spacecraft command upload on short timescales
is a necessity.

\acknowledgements I am very grateful to Vivek Dhawan (NRAO) for
sharing his re-reduction of the ground-based data from the HALCA
observation as part of a VLBA project on
\lsi~\citep{Dhawan:2006}. Many thanks to Russ Taylor, Bill Scott and
Marta Perecaula for reducing and analysing the data
originally published in \citet{Taylor:2000}. And lastly, to Ed
Fomalont and the VSOP team - many thanks for all the advice and help.



\end{document}